# A simple TEM method for fast thickness characterization of suspended graphene flakes

Sultan Akhtar[1, 2 *], Stefano Rubino[1*] and Klaus Leifer[1]

[1]*Department of Engineering Sciences, Uppsala University, Box 534, SE-751 21 Uppsala, Sweden*
[2]*Centre for Advanced Studies in Physics, Govt. College University, Katchery Road Lahore-54000, Pakistan*
* These authors have contributed equally to this work.

*Corresponding author; email:* <u>stefanorubino@yahoo.it</u> *(SR)*

**Abstract**

We present a simple and fast method for thickness characterization of suspended graphene flakes that is based on transmission electron microscopy (TEM) techniques. For this method, the dynamical theory of electron diffraction (Bloch-wave approach in two-beam case approximation) was used to obtain an analytical expression for the intensity of the transmitted electron beam $I_0(t)$, as function of the specimen thickness $t$ for thin samples ($t << \lambda$; where $\lambda$ is the absorption constant for graphite). We show that in thin graphite crystals the transmitted intensity is a linear function of the thickness. To obtain a more quantitative description of $I_0(t)$, high resolution (HR) TEM simulations are performed using the Bloch wave approach of the JEMS software [21]. From such calculations, we obtain $\lambda$ for a 001 zone axis orientation, in a two-beam case and in a low symmetry orientation. Subsequently, HR (used to determine $t$) and bright-field (to measure $I_0(0)$ and $I_0(t)$) images were acquired to experimentally determine $\lambda$. We obtain that the experimental value in the low symmetry orientation is close to the calculated value (i.e. *λ=225±9* nm for *300* kV accelerating voltage and 3 mrad collection angle). The simulations also show that the linear approximation obtained from the analytical expression is valid up to a sample thickness of several ten nanometres, depending on the orientation. When compared to standard techniques for thickness determination of graphene/graphite, the method we propose has the advantage of being relatively simple and fast, requiring only the acquisition of bright-field images.

**Introduction**

Graphene has several unique properties and the potential to replace many materials in a number of applications (sensors, supports, nanochips). This means that, in future, a large amount of graphene may be needed to fulfil the requirement of the World. Since the discovery of graphene in 2004 by Andre Geim, several production methods have been reported for single-layer graphene, for example [1-6], micromechanical cleavage of graphite, chemical vapour deposition, epitaxial growth, and oxidation of graphite. These methods have very low yield and use expensive starting materials for the production of graphene. Cheap and large scale production methods of graphene are still under development, but the need for a fast characterization method for graphene flakes has already arisen.

Ultrasound assisted graphene (UAG) exfoliation is a cheap and high-yield solution-based approach to the synthesis of large quantities of graphene that is currently being developed [5]. In this method, an entire graphite crystal is broken up into a large number of graphene-like flakes dispersed in a solution. A thickness characterization is needed in order to determine what percentage of the flakes is single-layer or few-layer graphene and therefore optimize the production parameters toward high–yield graphene synthesis. Among the standard techniques [1, 7-11] for thickness determination of a flake we mention: selected area electron diffraction (SAED), high-resolution transmission electron microscopy (HR-TEM), atomic force microscopy (AFM), Raman spectroscopy (RS), light optical microscopy (LOM) and Rayleigh scattering microscopy (RSM). They each have some drawbacks and difficulties in their application; therefore it is common to make use of more than one for the measurement of a specific flake. For instance, HR-TEM is the most direct identification tool; however, it can be very time-consuming [8, 11], and requires the presence of folded edges in the flake. It is possible to observe graphene layers in the LOM by placing them on top of oxidized Si substrates with typically 300 nm $SiO_2$, but the obtained thickness determination is only approximate. AFM has a very low throughput; moreover, the interaction forces between the AFM probe, the graphene flake, and the $SiO_2$ substrate lead to a measured thickness of 0.5-1 nm even for a single layer [6, 7, 12], which is bigger than what is expected from the interlayer graphite spacing. Thus, in practice, it is only possible to distinguish between one and two layers by AFM if the graphene films contain folds or wrinkles. Raman Spectroscopy (RS) can be used to identify single–layer or double-layer graphene by the presence of characteristic peaks, but it has a low spatial resolution [9]. Despite the wide use of micromechanical cleavage, the identification and counting of graphene layers is still a major hurdle. Mono-layers are a minority among accompanying thicker flakes and are hardly visible on most of the substrates under LOM [8, 13]. The same holds true for UAG, but the flakes thus obtained can then be easily captured on a holey carbon TEM grid for rapid inspection in the TEM. The method we present here provides thus a fast screening method to identify few-layer graphene flakes for further use.

In the graphite crystal, layers of carbon atoms are arranged in a hexagonal lattice [6, 8, 14, 15] forming graphite c-planes. The distance of these planes (002) in a graphite crystal is around 0.335 nm [12, 16, 17]. When a flake is folded, those planes are parallel to the electron beam and visible as dark lines with 0.335 nm spacings between them. By counting those lines in HR-TEM mode the thickness of the flake can be determined quite exactly. When compared to standard techniques for thickness determination of graphene, the method we propose has the advantage of being relatively simple and fast, requiring only the acquisition of bright-field (BF) images. The technique consists in measuring the reduction in the transmitted beam intensity which, in the approximations used here, decays linearly as function of the flake thickness. The method we propose has also equally the advantage of being local, *i.e.* it is not restricted to flakes of uniform thickness but can also be applied to regions where the graphene is folded.

**Theory and simulations**

In the TEM the incoming electron beam interacts with the specimen to form an image, opportunely magnified by the electromagnetic lenses. In the case of crystalline specimen, the incident beam is scattered into one direct (transmitted) beam and several diffracted beams (reflections, *g*). The dynamical theory of electron diffraction (Bloch-wave approach in two-beam case approximation) was used to obtain the following analytical expression for the intensity of the transmitted electron beam *T* (normalized to the incident intensity), as a function of specimen thickness *t* when considering absorption [18].

$$T = \frac{e^{(-2\pi t/\xi_0')}}{2(1+w^2)} \left[ \begin{array}{l} (1+2w^2)\cosh(\dfrac{2\pi}{\xi_g'\sqrt{1+w^2}}t) \\ +2w\sqrt{1+w^2}\sinh(\dfrac{2\pi}{\xi_g'\sqrt{1+w^2}}t) + \cos(\dfrac{2\pi\sqrt{1+w^2}}{\xi_g}t) \end{array} \right]. \quad (1)$$

Where, $\xi_0'$ is the mean absorption distance and $\xi_g'$ are the anomalous absorption distances for each reflection g and $\xi_g$ are the extinction distances. Their values were obtained by tables in the JEMS software [19]. The relation T + R= 1 (where R is the intensity of the reflected beams) is no longer valid because of absorption. It means that the amplitude of the Pendellösung oscillations decreases with increasing thickness [18]. We can rewrite Eq. (1) for exact Bragg condition *w = 0* as

$$T = \frac{1}{4}\left[ e^{-\left(\frac{2\pi}{\xi_0'} - \frac{2\pi}{\xi_g'}\right)t} + e^{-\left(\frac{2\pi}{\xi_0'} + \frac{2\pi}{\xi_g'}\right)t} + 2e^{-\left(\frac{2\pi}{\xi_0'}\right)t}\cos(\frac{2\pi}{\xi_g}t) \right]. \quad (2)$$

For thin specimens we can use the Taylor expansion for the exponential and cosine functions and neglect the quadratic and higher power terms, obtaining the following expression,

$$T = \frac{1}{4}\left[ (1-(\frac{2\pi}{\xi_0'}-\frac{2\pi}{\xi_g'})t) + \left(1-(\frac{2\pi}{\xi_0'}+\frac{2\pi}{\xi_g'})t\right) + 2\left(1-(\frac{2\pi}{\xi_0'}t)\right) \right]. \quad (3)$$

*T* is the change in the intensity of the direct beam after passing through the sample of thickness *t*, i.e. $T = I_0(t) / I_0(0)$. When $\xi_0' << \xi_g'$ we can neglect the terms containing $\xi_g'$ and write:

$$T = \frac{I_0(t)}{I_0(0)} = \frac{1}{4}\left[4 - 4(\frac{2\pi}{\xi_0'}t)\right] = \left(1 - \frac{2\pi}{\xi_0'}t\right). \quad (4)$$

Here we can define $\lambda = \xi_0'/2\pi$, as the material dependent absorption constant for electrons. Finally, we can write the following expression for the thickness of UAG flakes.

$$\frac{I_0(t)}{I_0(0)} = \left[1 - \frac{t}{\lambda}\right]. \tag{5}$$

The values for the intensity of the transmitted (000) beam $I_0(t)$ are obtained directly from bright-field images of UAG flakes. Eq. (5) is used to find the value of the absorption constant $\lambda$ for graphite by measuring $I_0(0)$ and $I_0(t)$ on a sample of known thickness $t$. UAG flakes were used for this purpose as well, since the thicknesses of the flakes can be measured through HR-TEM imaging of folded flake regions. Once $\lambda$ is known, the method can be readily used to obtain thickness maps directly from bright-field (BF) images.

An analytical formula can only be obtained for the so-called two-beam case. When more than just one reflection shows a significant intensity, this approach is no longer valid. We used the simulation package JEMS [19] to tackle the many-beam problem with the Bloch-wave approach. With this program, simulated HR images of a given crystal can be obtained for different orientations, thicknesses, collection angles and other acquisition parameters. However, it is not meaningful to try to obtain HR images in BF conditions (the aim of our technique), therefore we merely plot here the intensity of the 000 beam as $I_0(t)$. The Objective Aperture used in the experiment to obtain BF images is about 3 mrad in size, which means that, at 300 kV accelerating voltage, all reflections except (002) can be efficiently blocked. Since most flakes are in or close to a [001] zone axis (ZA) orientation, this effectively means blocking all beams except the 000, therefore the intensity in the final BF image can be considered to be simply the intensity of the 000 beam. We have performed simulations for different orientations, including cases where many $g$ reflections were excited (high symmetry orientation, close to the ZA). The results are shown in **Fig. 3** together with the experiment. It can be seen that the intensity decay (open circles) for the high symmetry orientation is much faster than for the low symmetry axis (solid triangles). The absorption constant $\lambda$ can be calculated from the slope of the curves, obtaining a value $\lambda_{sim}=208$ nm for the low symmetry orientation. The simulations also show that the linear approximation obtained from the analytical expression is valid up to several ten nm.

**Experiment**

The theory shows that the transmitted intensity is linearly dependent on the absorption constants. The simulation results for both high and low symmetry orientations are shown in **Fig. 4**. By considering these results we acquired BF images of UAG flakes at low symmetry orientation using SAED patterns. **Fig. 1** shows the sample and SAED patterns both at high and low symmetry orientations. The schematic diagram (panel a and b) shows how the apparent thickness is affected by sample orientation with respect to electron beam. The sample thickness and exit point of electron beam is indicated by t and P, respectively. The SAED patterns have same intensity for all spots once the crystal is exactly at zone axis (ZA). The SAED patterns show that the graphite crystal is close to (panel c) and away (panel d) from the ZA 001. The diffraction patterns in panel e and f are obtained from JEMS [19] for graphite at [001] and 7° away from the ZA, respectively. All the BF images of UAG flakes were taken at low symmetry orientation conditions confirming by SAED patterns.

For the TEM experiments UAG flakes [5] of various thicknesses and sizes were used to test the method for a variety of conditions. The flakes were collected directly from the solution by dipping commercially available TEM copper grids with amorphous holey carbon support films. The flakes are typically several μm across and less than 100 nm thick; they consist of one or more graphitic *c*-planes.

The selected area electron diffraction (SAED) patterns, the bright-field and high-resolution images were taken with a FEI Tecnai F30 TEM (300 kV).

**Results and Discussion**

As mentioned above, the folded edge of a graphene-like flake provides a clear TEM signature of the number of layers and therefore of the thickness. When the c-planes of a folded graphene sheet are locally parallel to the electron beam, dark lines spaced 0.335 nm appear, one for each layer [16], as is the case for TEM images of multi-walled carbon nanotubes. The folded edge of a bi-layer graphene exhibits two dark lines [1, 16], like double-walled nanotubes and a monolayer graphene shows only one dark line, like single-walled nanotubes. The drawing in **Fig. 2f** describes these observations. **Fig. 2a** is an image of one UAG flake suspended over a hole in the holey carbon film; the folding edge of the flake is indicated by a black arrow. About 17-18 layers can be counted in the HR image (panel b) or estimated from the intensity profile shown in panel d. The length of the folded part is estimated to be ~ 6.135 nm. The mean thickness, 12.27 ± 0.19 nm was calculated by measuring the thickness from more than six different parts of same edge in HR image. The plus/minus value is a *3σ*; where σ is an uncertainty in measurements calculated by standard deviation. The SAED pattern in panel c shows that the flake is about 5-7 degrees away from the 001 zone axis orientation and close to a two-beam case. These are the conditions under which **Eq. 5** was derived.

The values of the transmitted intensities $I_0(0)$ and $I_0(t)$ can be measured from a set of two BF images, one of the sample (yielding $I_0(t)$) and one of the vacuum ($I_0(0)$). A single BF image of the sample can also be used, provided that there are regions of vacuum in the image (i.e. holes in the carbon films), from which the value of $I_0(0)$ can be measured. The second method is simpler, but assumes that the recorded illumination is the same for every pixel, which may not be the case if, for example, the CCD gain correction is not properly performed or if the illumination is not uniform. In **Fig. 2**, the areas where we measured $I_0(t)$ from the flake is indicated by a white arrow in panel a. By inserting the measured values of intensities and thickness in **Eq. 5**, we calculated the absorption constant *λ=225±8 nm*. Where error is a total uncertainty in measurements (caused by intensities and thickness taken of *3σ* values) calculated by error propagation of statistics [20]. We applied the same method to measure *λ* from eight different flakes with a combination of SAED patterns, BF, HR images, obtaining a mean value of $\lambda_{exp}$~*225±9 nm*. This error is a standard deviation that describes the scatter of measurements about the average. The standard error or standard deviation of the mean was estimated to *±3* that is an estimation of standard deviation distribution of means. The data is given in **Table 1** and plotted along with simulation in **Fig. 3** below.

**Table 1**. *Calculations for scattering absorption constant, λ using Eq. 5*

| Intensity at vacuum $I_0(0)$ (e/ pixel) | Intensity at flake $I_0(t)$ (e/ pixel) | t (nm) | λ (nm) | Error propagation (nm) | Mean λ (nm) |
|---|---|---|---|---|---|
| 9015 | 8869 | 3.68 | 227 | 14 | 225±9 |
| 8980 | 8705 | 6.61 | 216 | 15 | |
| 8269 | 8019 | 7.15 | 236 | 11 | |
| 9323 | 8855 | 10.88 | 217 | 10 | |
| 8737 | 8261 | 12.27 | 225 | 8 | |
| 9259 | 8730 | 13.54 | 237 | 12 | |
| 9287 | 8534 | 18.80 | 232 | 10 | |
| 9764 | 8529 | 27,05 | 214 | 9 | |

Once λ for graphene-like flakes is determined, thickness maps of the samples can be constructed from a single or two BF images by using **Eq. (5)**. The thickness of a large number of flakes can be estimated rather quickly by taking images at medium magnification, without the need for time-consuming HR imaging. This also reduces the electron dose for each flake and consequently reduces the effects of beam damage. In **Fig. 4** we show a BF TEM image of one UAG flake (panel a) and the corresponding thickness map in nanometre units (panel b). The inset shows that the average thickness of the indicated area is ~4 nm.

**Sensitivity and detection limits**

The main source of error in this method originates from the Poisson's noise of the BF images:

$$\sigma_I = \sqrt{I_0(t)}; \qquad (6)$$

where $I_0(t)$ is the detected number of electrons per pixel (e/pixel) of the CCD camera. Since the camera saturates at about 2 500 e/pixel, the acquisition conditions are set to have a maximum value of about 1 500 e/pixel. This would give an error for a single pixel of ~39 e/pixel, which, combined with **Eq. 5**, gives a detection limit of ~6 nm (17 c-layers). However, assuming that a graphene flake has uniform thickness over a sufficiently large number of pixel (which can be achieved by choosing the appropriate magnification), one can reduce the error in the measurement of the intensity by taking the value of each pixel as an independent measurement of the intensity and obtain the Poissonian distribution of the signal. In our case of large numbers, this can be approximated with a Gaussian. The measured value of intensity would be the average over all the $N$ pixels for which the graphene flake has the same thickness; the error on this estimate would be

$$\sigma_{\langle I \rangle} = \sqrt{I_0(t)/N}. \qquad (7)$$

From this we can calculate the number of pixels needed to be able to resolve a single graphene layer (0.335 nm): from **Eq. 5** we obtain that a single graphene flake would cause the transmitted intensity to decrease by only 2 e/pixel and if we impose that the error is smaller than this value we need $N>375$ (or $N>3\,375$ for a 99.7% confidence).

**Conclusions**

A simple TEM method was developed and applied for fast thickness characterization of suspended graphene flakes. The method is based on the assumption that the intensity of the transmitted beam $I_0(t)$ is a linear function of the thickness. We have shown that this is valid for thin graphene-like flakes close to a [001] orientation but at least 5° away from it. This is a peculiarity of graphite, for which the characteristic length for absorption (which follow an exponential law) is smaller than for diffraction (which is quadratic). Therefore, when only a few diffracted beams are excited, as it is the case in low symmetry orientations, pure absorption describes best the intensity variation as function of thickness, i.e. the quadratic terms can be neglected in favour of the linear one from the exponential decay. However when many beams are excited in a high symmetry orientation one has that the sum of the smaller quadratic terms outweighs the linear term and the intensity variation deviates from the linear law and exhibits a stronger oscillatory decay.

When the conditions for linear decay are fulfilled, $I_0(t)$ can be measured by acquiring BF images at medium resolution with an objective aperture small enough to block all beams except (000), which means the aperture should block electron scattering to angles equal to and larger than the (100)-type reflections. We have determined that the characteristic absorption length in graphene is $\lambda = 225 \pm 9$ nm. From the BF images one can construct the thickness maps of each flake, making thus possible to determine the thickness even for folded or wrinkled flakes. When compared to conventional HR TEM imaging, the method is much faster, which had the twofold advantage of enabling fast screening of several dozens of flakes in a single TEM session and of greatly reducing the electron dose on each flake and thus beam damage. When compared to SAED it has a higher spatial resolution and it is rather unaffected by local variations of orientations (wrinkles, folding), provided they are not too close to the [001] zone axis. The method is similar to thickness maps obtained by EFTEM techniques [21], but it does not require an energy filter and the resulting images have higher intensity and therefore lower relative noise levels.

**Figure captions**

**Figure 1. Description of graphite crystal orientations:** *Schematic of high (a) and low (b) symmetry orientations where P is the exit point of the electron beam and t the thickness of the sample; the experimental SAED patterns close to a high (c) and low (d) symmetry axis i.e. close to and away from zone axis 001 respectively; diffraction patterns of graphite simulated by JEMS at high (e) and low (f) symmetry orientations.*

**Figure 1. TEM-thickness measurements of graphene-like flakes:** *a) Bright-field TEM image of one folded UAG flake. The area 2 (indicated by white arrow) was used to measure intensity $I_0$ (t) from the flake and vacuum for $I_0$ (0). The inset is a selected area electron diffraction (SAED) pattern at the orientation away from the zone axis (ZA) taken from the unfolded part of the flake (area indicated by white arrow head); b) High resolution image of the folded edge of the flake taken from the indicated area in panel a (black arrow). The flake is 17-18 layers thick here, as indicated by the intensity profile in panel c; c) The intensity profile shows that the number of layers in the marked region can be estimated to 17-18 or ~6.135 nm; d) The drawing of the folded edge of the flake shows how c-planes become parallel to the incident beam of electrons and appear in the HR image. The thickness of the folded edge was measured at area 1 (indicated by a pair of black arrows) and it is clear that the thickness at area 2 (indicated by a pair of red arrows) will be double of area 1.*

**Figure 3. Simulation and experimental results of graphite:** *the experimental values (black solid circles) obtained by measuring transmitted intensities and thickness from BF and HRTEM images; the simulations (open circles/solid triangles) are obtained using the JEMS software in a high/low symmetry orientations; the cross/plus signs curve is obtained by using the analytical expression for transmitted intensity derived for the two-beam case approximation at close/away from ZA. The slope of the curves gives the value of the absorption constant λ for graphite.*

**Figure 4.** *a) BF TEM image of one graphene-like flake (indicated by arrow); b) the corresponding thickness map obtained by applying the TEM method proposed in this work. The flake is estimated to be about 4 nm thick in the region shown in the intensity profile (inset).*

**Figure 1**

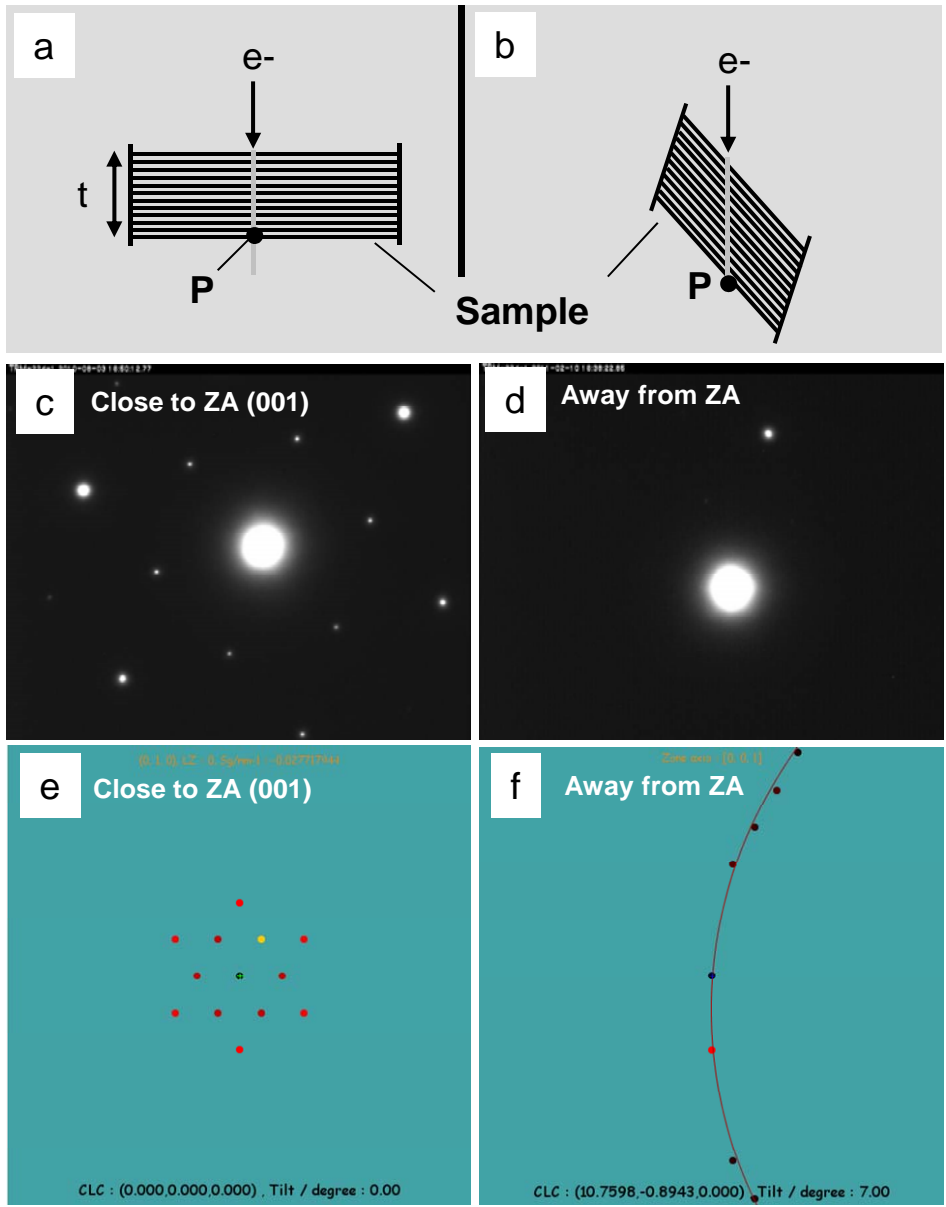

**Figure 2**

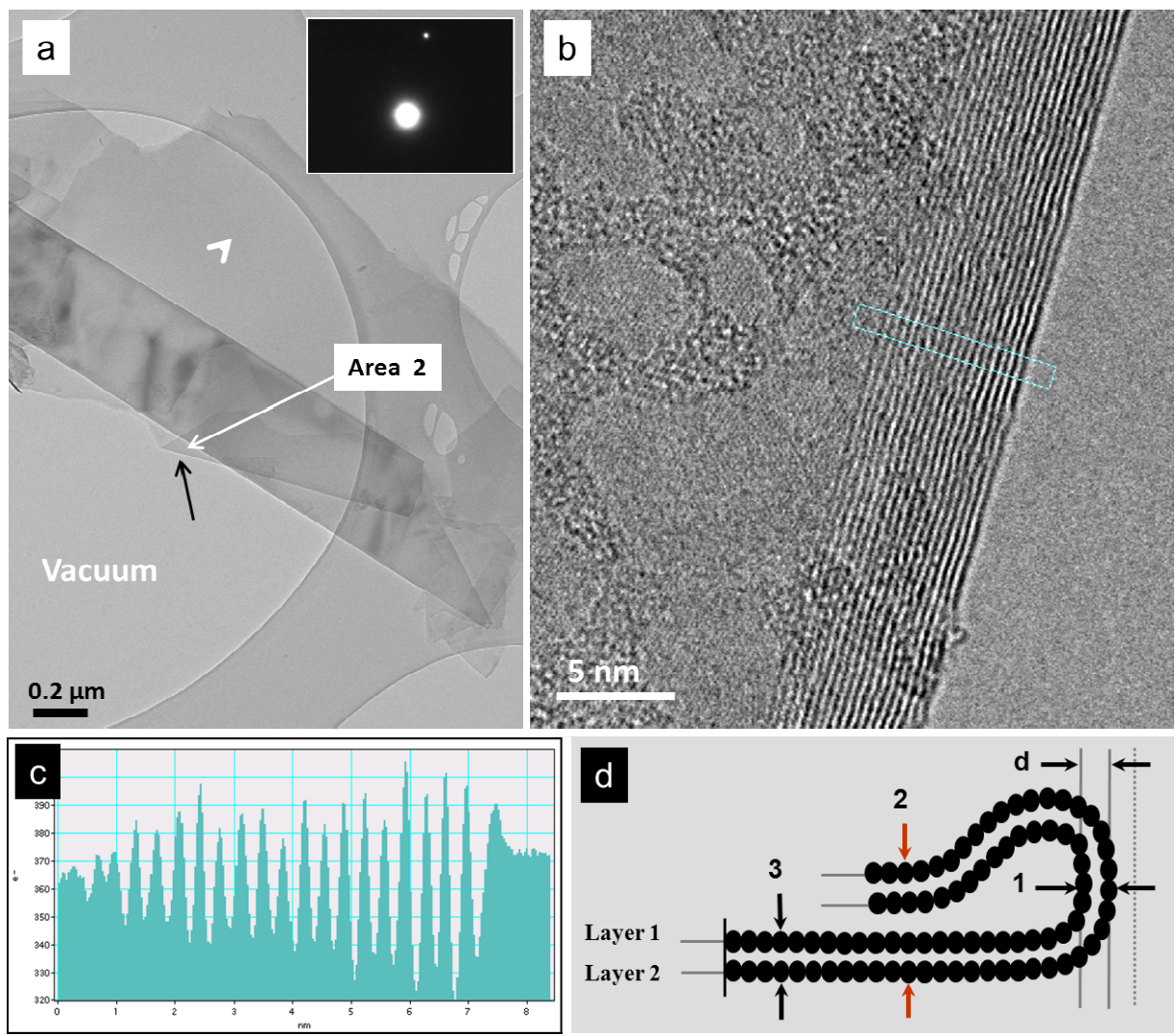

**Figure 3**

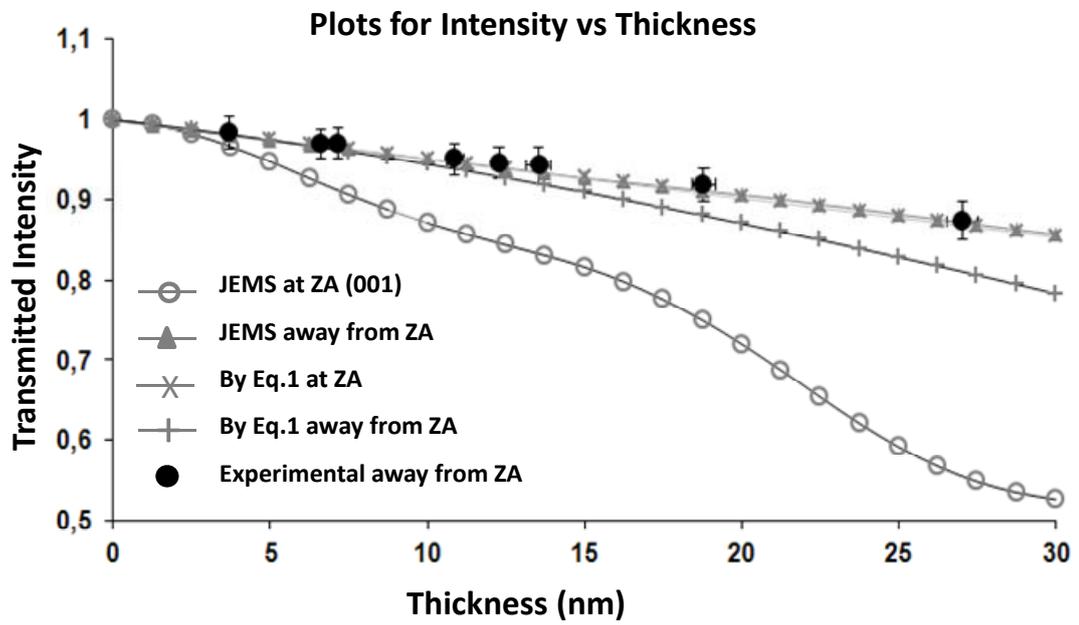

**Figure 4**

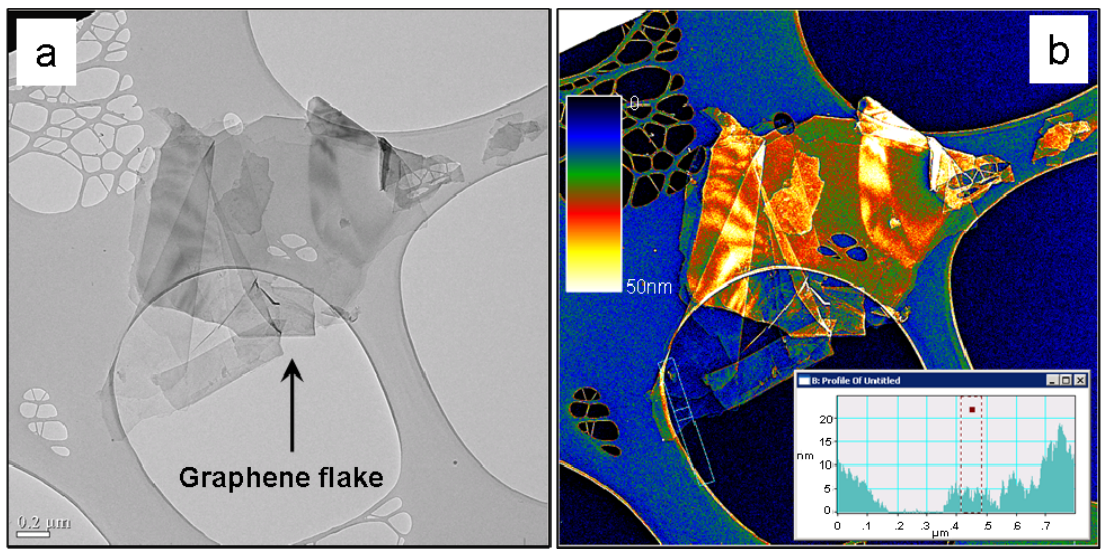